\documentclass[12pt,draftclsnofoot,onecolumn]{IEEEtran}
\usepackage{graphicx}
\usepackage{caption}
\ifCLASSOPTIONcompsoc
\usepackage[caption=false,font=normalsize,labelfont=sf,textfont=sf]{subfig}
\else
\usepackage[caption=false,font=footnotesize]{subfig}
\fi
\usepackage{indentfirst}

\usepackage{amsmath}
\allowdisplaybreaks[4]
\usepackage{amssymb}
\usepackage{times}
\usepackage{mathtools}
\usepackage{psfrag}
\usepackage{cite}
\usepackage{lastpage}
\usepackage{fancyhdr}
\usepackage{color}
\usepackage{amsthm}
\usepackage{bigints}
\newtheorem{Theorem}{Theorem}
\newtheorem{Lemma}{Lemma}

\newtheorem{Corollary}{Corollary}

\newtheorem{lemma}[Lemma]{$\mathbf{Lemma}$}

\begin{document}
\title{Hybrid Successive Interference Cancellation and Power Adaptation: a Win-Win Strategy for Robust Uplink NOMA Transmission}
\author{Yanshi Sun, \IEEEmembership{Member, IEEE}, Wei Cao, Momiao Zhou, \IEEEmembership{Member, IEEE}, Zhiguo Ding, \IEEEmembership{Fellow, IEEE}
\thanks{
Y. Sun, Wei Cao and M. Zhou  are  with the School of Computer Science and Information
Engineering, Hefei University of Technology, Hefei, 230009, China. (email: sys@hfut.edu.cn, caowei0115@163.com and mmzhou@hfut.edu.cn).

Z. Ding is with Department of Electrical Engineering and Computer
Science, Khalifa University, Abu Dhabi, UAE, and Department of Electrical
and Electronic Engineering, University of Manchester, Manchester, UK. (email: zhiguo.ding@manchester.ac.uk).
}\vspace{-3em}}
\maketitle
\begin{abstract}
The aim of this paper is to reveal the importance of hybrid successive interference cancellation (SIC) and power adaptation (PA) for improving transmission robustness of uplink non-orthogonal multiple access (NOMA).
Particularly, a cognitive radio inspired uplink NOMA communication scenario is considered, where one primary user is allocated one dedicated resource block, while $M$ secondary users compete with each other to be opportunistically served by using the same resource block of the primary user. Two novel schemes are proposed for the considered scenario, namely hybrid SIC with PA (HSIC-PA) scheme and fixed SIC with PA (FSIC-PA) scheme. Both schemes can ensure that the secondary users are served without degrading the transmission reliability of the primary user compared to conventional orthogonal multiple access (OMA) based schemes. Rigorous analytical results are presented to evaluate the performance of the proposed two schemes. It is shown that both schemes can avoid outage probability error floors without any constraints on users' target rates in the high SNR regime. Furthermore, it is shown that the diversity gain achieved by the HSIC-PA scheme is $M$, while that of the FISC-PA scheme is only $1$. Numerical results are provided to verify the developed analytical results and also demonstrate the superior performance achieved by the proposed schemes by comparing with the existing HSIC without PA (HSIC-NPA) scheme.
The presented simulation results also show that HSIC-PA scheme performs the best among the three schemes, which indicates the importance of the combination of HSIC and PA for improving transmission robustness.
\end{abstract}
\begin{IEEEkeywords}
Non-orthogonal multiple access (NOMA), hybrid successive interference cancellation (HSIC), power adaptation, outage probability.
\end{IEEEkeywords}
\section{Introduction}
Non-orthogonal multiple access (NOMA) has attracted extensive research interest during the past few years, and has been recognized as an important potential enabling technology for future wireless communication systems \cite{3GPPNOMAR13,3GPPNOMAR16,you2021towards}. Compared to conventional orthogonal multiple access (OMA), where one channel resource block can be accessed by a single user only, the key appealing feature of NOMA is that allowing multiple users to simultaneously access the same channel resource block is encouraged \cite{makki2020survey,saito2013non, ding2014performance}. Thus, by applying NOMA, larger connectivity and higher spectral efficiency can be obtained.
Existing research works  show that NOMA can be compatible with many other advanced technologies, such as multiple input multiple output (MIMO) \cite{choi2016power,chen2016application,zhu2020optimal}, millimeter wave communications \cite{sysmmwave2018}, Terahertz communications \cite{zhang2020energy,ding2023joint,zuo2023non}, reconfigurable intelligent surfaces (RIS) \cite{wu2021coverage,li2023achievable,ding2020impact}, satellite communications \cite{lin2019joint,gao2021sum,liu2019qos} and so on.

Since NOMA allows multiple users to simultaneously occupy one channel resource block, how to address inter-user interference is one of key issues in NOMA communication systems. To this end, a widely used method in NOMA to address inter-user interference is successive interference cancellation (SIC), where users' signals are decoded in a successive manner \cite{higuchi2013non}. Due to the error propagation nature of SIC, how to order users plays a very important role in the performance of SIC. Conventionally, there are two main types of methods for determining  the decoding order of users in NOMA. One  is known as the channel state information (CSI) based SIC method, where users are ordered according to the quality of their channels \cite{higuchi2013non, gao2017theoretical,Xia2018outage}. The other is known as the quality of service (QoS) based SIC method, where the signals for the users with more stringent QoS are decoded first, while other users are often opportunistically served and their signals are decoded later \cite{zhou2018state,Dhakal2019noma,Ding2019simple}. Note that, most existing works on NOMA carried out a prefixed SIC decoding order according to either the above two aforementioned criteria. Unfortunately, a very dispiriting phenomenon exists in the NOMA schemes based on the aforementioned CSI or QoS based methods. Specifically, the outage probability achieved by these schemes suffers from severe error floors, which means that the outage probability achieved by
a certain user doesn't approach zero as SNR goes infinity. Thus, the transmission reliability cannot be guaranteed, which significantly limits the application of NOMA in many practical scenarios.
It was thought that,  such outage probability error floors are unavoidable in the implementation of NOMA, and swapping SIC decoding orders dynamically cannot yield a significant performance gain \cite{ding2021new,ding2020unveiling1,ding2020unveiling2,lu2022advanced}.

Motivated by the error floor issue, a new design of SIC namely hybrid SIC (HSIC) was initially proposed for cognitive radio inspired uplink NOMA by \cite{ding2021new,ding2020unveiling1,ding2020unveiling2}. In the proposed HSIC scheme, the decoding orders of users are dynamically determined according to the relationship between the instantaneous channel conditions and users' target rates. \cite{ding2021new,ding2020unveiling1,ding2020unveiling2} show that the proposed HSIC scheme can avoid outage probability error floors, under some constraints on users' target rates. The most important contributions of the series studies in \cite{ding2021new,ding2020unveiling1,ding2020unveiling2} are two folds.
First, \cite{ding2021new,ding2020unveiling1,ding2020unveiling2} showed that it is possible to avoid outage error floors, at least under some specific conditions. Second, \cite{ding2021new,ding2020unveiling1,ding2020unveiling2} indicated the importance of introducing HSIC to improve transmission robustness of NOMA.

However, as mentioned above, the proposed scheme in \cite{ding2021new,ding2020unveiling1,ding2020unveiling2}  can only avoid outage probability error floors under some stringent conditions on users' target rates, which may not be met in many realistic scenarios. Thus, it is natural to ask the following two questions.
The first question is whether it is possible to avoid outage probability error floors without any constraints on users' rates. And the second question is whether it is necessary to apply HSIC to avoid outage probability error floors.

This paper aims to answer the two aforementioned questions, and investigate the impact of the combination of HSIC and power adaptation (PA) on improving the transmission robustness in NOMA. Specifically,  a cognitive radio inspired uplink NOMA scenario is considered. In the considered scenario, one primary user is allocated one dedicated channel resource block, while there are $M$ secondary users who compete with each other to opportunistically share the primary user's resource block without degrading the outage performance of the primary user. Two new designs of NOMA schemes, namely HSIC with PA (HSIC-PA) and fixed SIC with PA (FSIC-PA) are proposed. Both schemes can avoid outage probability error floors without any constraints on users' target rates. The main contributions of this paper are listed as follows.

\begin{itemize}
  \item Two novel designs of uplink NOMA schemes are proposed, namely HSIC-PA and FSIC-PA\footnote{Note that the
        HSIC-PA scheme extends the scheme proposed in our previous work \cite{sun2021new} where only two users are considered, while the FSIC-PA scheme hasn't been proposed according to our best knowledge.}. In the proposed HSIC-PA scheme, the decoding order of the secondary user can be dynamically adjusted according to the channel conditions. While in the proposed FSIC-PA scheme, the decoding order of the secondary user  is fixed at the  second stage of SIC. By rigorous derivation, the closed-form expressions for the outage probabilities  achieved by the proposed two schemes are obtained.
  \item  Based on the obtained expressions for the outage probabilities, asymptotic analysis in the high SNR regime is further developed to gain more insights into the proposed two schemes.  It is shown that both HSIC-PA scheme and FSIC-PA scheme can avoid outage probability error floors without any constraints on users' target rates. The fact that the proposed FSIC-PA scheme can avoid error floors indicates that HSIC is not necessary to avoid error floors. Furthermore, the diversity gains achieved the proposed two schemes are also provided, respectively. Interestingly, the diversity gain achieved by HSIC-PA scheme is $M$, whereas that achieved by FSIC-PA scheme is only $1$.
  \item Numerical results are presented to verify the accuracy of the developed analytical results and  demonstrate the superior performance of the proposed HSIC-PA scheme and FSIC-PA scheme, by comparing with the benchmark scheme termed HSIC-NPA proposed in \cite{ding2021new,ding2020unveiling1,ding2020unveiling2}. In terms of outage probability and ergodic rate, it is shown that FSIC-PA scheme performs better than HSIC-NPA scheme in the high SNR regime, but worse in the low SNR regime. Besides, HSIC-PA scheme performs the best among three schemes at all SNRs in terms of outage probability and ergodic rate, which shows the power of the combination of HSIC and PA in the design of uplink NOMA transmissions. In terms of power consumption,  both the proposed HSIC-PA and FSIC-PA schemes consume less power than the existing HSIC-NPA scheme, whereas HSIC-PA scheme is more power-consuming than FSIC-PA scheme.
\end{itemize}

\section{System Model}
	Consider an uplink NOMA communication scenario with one base station (BS), one primary user $U_0$ and $M$
	secondary users $U_m$, $1\le m\le M$. Note that, in the considered scenario, ensuring the transmission reliability of $U_0$ is of the high priority, which has a target data rate denoted by $R_0$. In conventional OMA based schemes, the primary user is allocated with a dedicated resource block, which cannot be accessed by other users. While in the considered NOMA schemes of this paper, $M$ secondary users compete with each other to opportunistically  access the channel resource block which is allocated to the primary user. Note that allowing secondary users to share the channel resource block of the primary user must be done in such a way to ensure that the QoS of the primary user $U_{0}$ is not degraded.
	
	The channel gain of the primary user $U_{0}$ is denoted by $g$, and the channel gains of the secondary users are denoted by $h_m$, $1\le m\le M$. In this paper, $g$ and $h_m$ are modeled as the normalized Rayleigh fading gains, which means that $g$ and $h_m$ are independent and identically distributed (i.i.d) circular symmetric complex Gaussian (CSCG) random variables with zero mean and unit variance, i.e., $g\sim\mathcal{CN}(0,1)$ and $h_m \sim \mathcal{CN}(0,1)$. The transmit power of the primary user $U_{0}$ is denoted by $P_{0}$. The transmit power of the secondary user $U_m$ is denoted by $\beta P_s$, where
	$\beta \in \left [0,1\right ]$ is the adjustable power adaptation coefficient of $U_{m}$, and $P_s$ is the maximum power of $U_m$. Without loss of generality, the background noise power is also assumed to be normalized  throughout the paper.
	
In the remainder of the paper, the $M$ secondary users are ordered according to their channel gains:
	\begin{align}\label{|h_m|^2}
		\left | h_{1} \right | ^2< \cdots <\left |  h_{M}\right|^2.
	\end{align}
	
	In this paper, two novel NOMA schemes are proposed, namely HSIC-PA scheme and FSIC-PA scheme.
	It will be shown that both schemes can avoid outage probability error floors.
	For each scheme, in each period of transmission, only the secondary user which can achieve the largest instantaneous achievable rate is allowed to transmit signal by sharing the primary user's resource block.
	The proposed two schemes are described in the following two subsections.
	
	\subsection{HSIC-PA Scheme}
	
	To begin with, define an interference threshold denoted by $\tau (g)$ as follows:
	\begin{equation}
		\begin{aligned}
			\tau (g)=&\max\left \{ 0,\frac{P_{ 0}\left |g  \right | ^2 }{2^{R_{ 0}} -1} -1\right \}.
		\end{aligned}
	\end{equation}
	Note that $\tau(g)$ can be interpreted as the maximum interference, with which $U_0$ can
	still achieve the same outage performance as in OMA where the resource block would be  solely occupied by $U_0$. For more details on $\tau(g)$, please refer to \cite{sun2021new,ding2021new}.
	{\color{black} Define  $\epsilon_{0}=2^{R_{0}}-1$ and $\alpha _{0}=\frac{\epsilon_{0}}{P_{0}}$, we have
	$$\tau(g)=\begin{cases}
		|g|^2\alpha_{0}^{-1}-1 ,&|g|^2>\alpha_{0},\\
		0 ,&|g|^2<\alpha_{0}.
	\end{cases}$$}
	
	For each secondary user $U_m$, its instantaneous achievable rate is determined by how its channel
	gain compares to $\tau (g)$, which can be classified into the following two types:
	\begin{itemize}
		\item Type I: the maximum received signal power of $U_m$ at the BS is less than or equal to $\tau(g)$, i.e.,
		$P_{s}\left | h_{m} \right |^2 \le \tau (g) $. For this case, putting $U_m$ at the second stage of SIC
		can yield a larger data rate compared to putting $U_m$ at the first stage of SIC, and will not prevent the primary user from successfully decoding its signal. Thus, it is favorable to decode $U_m$'s signal at the second stage of SIC, and the achievable rate of $U_m$ is given by
		\begin{align}
			R_{\uppercase\expandafter{\romannumeral1}}^m=\log(1+P_{ s}\left | h_{ m}  \right |^2),
		\end{align}
which is the same as in HSIC-NPA scheme proposed in \cite{ding2021new}.
		\item Type II: the maximum received signal power of $U_m$ at the BS is larger than $\tau (g)$, i.e., $P_{s}\left | h_{m} \right |^2 > \tau (g)$. For this case, the benchmark scheme termed HSIC-NPA which is proposed in \cite{ding2021new}
		only considers the case where $\beta$ is set to be $1$. Thus, in order to avoid degrading the QoS of $U_0$, $U_m$'s signal can only be decoded at the first stage of SIC in HSIC-NPA, yielding the following achievable data rate of $U_m$:
		\begin{align}
			R_{II,1}^m=\log(1+\frac{P_{s}\left | h_{m} \right |^2 }{P_{0}\left | g \right | ^2+1} ).
		\end{align}
		Note that the drawback of putting $U_m$ at the first stage of SIC is that, when $P_0|g|^2$ is large,  $R_{II,1}^m$ might still be small even with a large $P_{s}\left | h_{m} \right |^2$.
		To this end, the proposed HSIC-PA scheme offers an additional choice where $\beta$ can be set to be less than $1$ so that $\beta P_s|h_m|^2=\tau(g)$, which can provide an opportunity to yield a larger achievable rate. As a result, $U_m$'s signal can be decoded at the second stage of SIC, yielding the following achievable data rate of $U_m$:
		\begin{align}
			R_{\uppercase\expandafter{\romannumeral2},2}^m=\log(1+\tau(g)).
		\end{align}
		Thus, in the proposed HSIC-PA scheme, when $P_{s}\left | h_{m} \right |^2 > \tau(g)$, the achievable data rate of $U_m$ is given by:
		\begin{align}
			R_{\uppercase\expandafter{\romannumeral2}}^m=\max\left\{R_{\uppercase\expandafter{\romannumeral2},1}^m,R_{\uppercase\expandafter{\romannumeral2},2}^m\right\}.
		\end{align}
	\end{itemize}
	According to the above discussions, the achievable data rate of $U_m$ in HSIC-PA scheme can be concluded as:
	\begin{align}
		R^m=\begin{cases}
			R_{\uppercase\expandafter{\romannumeral1}}^m,&P_{s}\left | h_{ m}  \right |^2 \le\tau(g) \\
			R_{\uppercase\expandafter{\romannumeral2}}^m,&P_{s}\left | h_{ m}  \right |^2 >\tau(g).
		\end{cases}
	\end{align}
	\subsection{FSIC-PA Scheme}
	Another scheme termed FSIC-PA is proposed in this subsection.
	Note that in HSIC-PA scheme, the secondary user's signal can be decoded either at the first or second stage of SIC. However, in FSIC-PA scheme, its signal can only be decoded at the second stage of SIC.
	
	In FSIC-PA scheme, for each secondary user $U_m$, its instantaneous achievable rate can also be determined by considering the following two cases as in the previous subsection.
	\begin{itemize}
		\item Type I: the maximum received signal power of $U_m$ at the BS is less than or equal to $\tau(g)$, i.e.,
		$P_{s}\left | h_{m} \right |^2 \le \tau (g) $. For this case,  the decoding strategy is as same as in the HSIC-NPA and the proposed HSIC-PA scheme, where $U_m$ is decoded at the second stage of SIC. Thus, the achievable data rate of $U_{m}$ is $\hat{R}^{m}_I =\log(1+P_{s}|h_{m}|^2)$, since the interference from $U_0$ can be removed by SIC.
		\item Type II: the maximum received signal power of $U_m$ at the BS is larger than $\tau (g)$, i.e., $P_{s}\left | h_{m} \right |^2 > \tau (g)$. For this case, in the proposed FSIC-PA scheme, $U_m$ can only be decoded at the second stage of SIC. To carry out this strategy, $\beta$ is set to be less than $1$ so that $\beta P_s|h_m|^2=\tau(g)$. Thus, the achievable data rate of $U_m$ for type II is $\hat{R}_{II}^m=\log(1+\tau(g))$.
	\end{itemize}
	
	By concluding the above two cases, the achievable data rate of $U_m$ in the FSIC-PA scheme can be expressed as:
	\begin{align}
		\hat{R}^m=\begin{cases}
			\hat{R}^m_{I},&P_{s}\left | h_{ m}  \right |^2 \le\tau(g) \\
			\hat{R}^m_{II},&P_{s}\left | h_{ m}  \right |^2 >\tau(g).
		\end{cases}
	\end{align}
	
	Note that, the proposed HSIC-PA and FSIC-PA schemes can ensure that the outage performance of the primary user is the same as that in the OMA scheme.  Because the use of NOMA is transparent to the primary user,  this paper focuses on the performance of the opportunistically served secondary users.

\section{Performance Analysis on HSIC-PA scheme and FSIC-PA scheme}
In this section, the closed-form expressions for the outage probabilities of the served secondary user achieved by the proposed two schemes will be provided. Furthermore, asymptotic analysis for the outage probabilities will be presented, which shows that both HSIC-PA and FSIC-PA schemes can avoid outage probability error floors without any constraints on users' target rates. Besides, rigorous comparisons between the proposed HSIC-PA/FSIC-PA scheme with the existing HSIC-NPA scheme will be carried out.

\subsection{Outage probability achieved by HSIC-PA scheme}
This subsection provides the exact and asymptotic expressions for the overall outage probability
of the served secondary users achieved by the proposed HSIC-PA scheme. Besides, the diversity gain \cite{tse2005fundamentals} achieved by HSIC-PA is also provided.

 Assume that all the secondary users have the same target rate, denoted by $R_s$. The overall outage probability achieved by the served secondary users in HSIC-PA is given by:
	\begin{align}\label{P_out}
		P_{out}=\text{Pr}\left(\max\{R^m, 1\leq m\leq M\}<R_s\right).
	\end{align}

	For the ease of characterizing the outage probability $P_{out}$, it is helpful to define the event $E_m$, which denotes the event that there are $m$ secondary users belonging to type I. Particularly, $E_m$ can be expressed as follows:
	\begin{align}\label{E_{m}}
		E_{m}\!=\!\begin{cases}
			\left \{ \left |h_{m} \right |^2< \frac{\tau (g)}{P_{s}},\left | h_{m+1} \right | ^2>\frac{\tau (g)}{P_{s}}  \right \},
			&1\le m\le M-1,\\
			\left\{|h_{1}|^2 > \frac{\tau (g)}{P_{s}}\right\}, & m=0, \\
			\left\{|h_{M}|^2 < \frac{\tau (g)}{P_{s}}\right\}, & m=M,
		\end{cases}
	\end{align}
	where the extreme cases $E_0$ and $E_{M}$ denote the events where there is no type I secondary users and all the secondary users belong to type I, respectively.
	
It is shown that the expression  of $P_{out}$ can be divided into four parts, as highlighted in the following lemma.

\begin{lemma}
  For ease of calculation, $P_{out}$ can be further simplified as:
	\begin{align}\label{A}
		P_{out}=&\underbrace{P\left(|h_M|^2>\frac{\tau(g)}{P_s},\left | h_{M} \right |^2<\left |\overline{h_{k}}\right |^2,R^M_{II,2}<R_s,|g|^{2}>\alpha_{0} \right)}_{\tilde{Q}_1} \notag\\
		+&\underbrace{P\left(|h_M|^2>\frac{\tau(g)}{P_s},\left | h_{M} \right |^2>\left |\overline{h_{k}}\right |^2,R^M_{II,1}<R_s,|g|^{2}>\alpha_{0} \right)}_{\tilde{Q}_2}\notag \\
		+&\underbrace{P\left( E_{M},R^M_I<R_{s},\left | g \right | ^2>\alpha _{0}  \right)}_{Q_{M}}  +\underbrace{ P\left(
			 R^M_{II}<R_{s} ,|g|^2<\alpha_{0} \right) }_{Q_{M+1}}.
	\end{align}
\end{lemma}
\begin{IEEEproof}
	Please refer to Appendix A.	
\end{IEEEproof}

By deriving the expressions of $\tilde{Q}_1$, $\tilde{Q}_2$, $Q_{M}$ and $Q_{M+1}$ as shown in Appendix B, the expression for  the overall outage probability of the admitted secondary users in HSIC-PA scheme can be obtained as shown in the following theorem.
\begin{Theorem}
	The overall outage probability $P_{out}$ of the admitted secondary users in HSIC-PA can be expressed as follows:
\begin{align}\label{T1}
{\color{black}P_{out}=\sum\limits_{i=0}^{M}\left(\begin{array}{c}
		M \\
		i
	\end{array}\right)(-1)^ie^{-i\alpha_{s}}\frac{1-e^{-(\alpha_{s}P_{0}i+1)\alpha_{1}}}{\alpha_{s}P_{0}i+1}+(1-e^{-\alpha_{s}})^{M}e^{-\alpha_{1}},}
\end{align}
	where $\epsilon_{s}=2^{R_{s}}-1$,
	$\alpha_{s}=\frac{\epsilon_{s}}{P_{s}}$,
	$\alpha_{1}=(1+\epsilon_{s})\alpha_{0}$.	
\end{Theorem}
\begin{IEEEproof}
	Please refer to Appendix B.
\end{IEEEproof}

Based on Theorem $1$, the asymptotic expression for $P_{out}$ in the high SNR regime can be obtained as shown in the following corollary.

\begin{Corollary}
At high SNR, i.e., $P_0=P_s\rightarrow \infty$, the overall outage probability of the served secondary users in HSIC-PA can be approximated as follows:
\begin{align}\label{HPout}
	{\color{black}	P_{out}\approx \frac{\epsilon_{s}^M}{P_s^MP_0}\sum\limits_{i=1}^{M}\left(\begin{array}{c}
			M \\
			i
		\end{array}\right)\frac{\epsilon_{0}^{i+1}(1+\epsilon_{s})^{i+1}}{i+1}-\frac{\epsilon_{s}^M}{P_s^MP_0^2}\sum\limits_{i=0}^{M}\left(\begin{array}{c}
			M \\
			i
		\end{array}\right)\frac{\epsilon_{0}^{i+2}(1+\epsilon_{s})^{i+2}}{i+2}+\frac{\epsilon_{s}^M}{P_s^M}.}
\end{align}
\end{Corollary}
\begin{IEEEproof}
	Please refer to Appendix C.
\end{IEEEproof}

Further, it is straightforward that the first two terms of (\ref{HPout}) can be omitted in the high SNR regime, yielding a more simplified expression for $P_{out}$, as highlighted in the following corollary.

\begin{Corollary}
At high SNR, i.e., $P_0=P_s\rightarrow \infty$,
the approximation of $P_{out}$ shown in (\ref{HPout})  can be further approximated as follows:
\begin{align}\label{HCPout}
  P_{out}\approx\frac{\epsilon_{s}^M}{{P_s}^M}.
\end{align}
\end{Corollary}

\textbf{\textit{Remark 1.} }Note that, the existing HSIC-NPA scheme can only avoid outage probability error floors under the constraint that $\epsilon_{0}\epsilon_{s}\leq 1$, which means that the feasible target rate for reliable transmission of the secondary users is primarily restricted by that of the primary user.
However, from the results shown in Corollary $2$, it can be easily concluded that the outage probability error floor can be avoided by HSIC-PA scheme without any constraints on the users' target rates.  Hence, the first question raised in Section I can be answered with the answer that it is possible  to avoid outage probability error floors without any constraints on users' target rates.

\textbf{\textit{Remark 2.} }In wireless communications, diversity gain is usually used as an important performance metric to measure how fast the outage probability decreases as transmit power increases \cite{tse2005fundamentals}. It denotes the asymptotic scaling law of the outage probability to the transmit SNR. Specifically, the diversity gain, say $d$, achieved by HSIC-PA is defined as:
 \begin{align}
d=-\lim\limits_{P_s\rightarrow\infty}\frac{\log P_{out}}{\log P_s}
\end{align}

Based on the results shown in Corollary $2$, it can be straightforwardly obtained that
$d=M$. Therefore, the diversity gain achieved by the HSIC-PA scheme is $M$, which is exactly the number of the secondary users. Thus, multi-user diversity gain can be fully utilized by the proposed HSIC-PA scheme, which means increasing the number of secondary users is helpful to reduce the overall outage probability.

From the perspective of diversity gain, the difference between the HSIC-NPA scheme and the HSIC-PA scheme can also be revealed. Recall that the diversity gain achieved by HSIC-NPA is also $M$ when $\epsilon_{0}\epsilon_{s}\leq1$, otherwise a diversity gain of zero is realized.

\subsection{Outage probability achieved by FSIC-PA scheme}
This subsection provides the exact  expression for the overall outage probability
of the served secondary users in the proposed FSIC-PA scheme. Asymptotic analysis for the outage probability is also provided.

For the FSIC-PA scheme, the overall outage probability achieved by the served secondary users is defined as:
\begin{align}\label{hatP_out}
	\hat{P}_{out}=\text{Pr}\left(\max\{\hat{R}^m, 1\leq m\leq M\}<R_s\right).
\end{align}

The following theorem provides the closed-form expression for the outage probability achieved by the FSIC-PA scheme.

\begin{Theorem} The overall outage probability $\hat{P}_{out}$ of the served secondary users in  FSIC-PA  can be expressed as follows:
	\begin{align}\label{Pout_F1}
		\hat{P}_{out}=1-e^{-\alpha_{1}}+(1-e^{\alpha_{s}})^Me^{-\alpha_{1}}.
	\end{align}
\end{Theorem}
\begin{IEEEproof}
	Please refer to Appendix D.
\end{IEEEproof}

Based on Theorem $2$, asymptotic expression for $\hat{P}_{out}$ in the high SNR regime can be obtained as shown in the following corollary.

\begin{Corollary}
	At high SNR, i.e., $P_0=P_s\rightarrow \infty$, the overall outage probability of the served secondary users in the FSIC-PA scheme can be approximated as follows:
	\begin{align}
		\hat{P}_{out}\approx \frac{\epsilon_{0}(1+\epsilon_{s})}{P_0}+\frac{\epsilon_{s}^M}{P_s^M}-\frac{\epsilon_{s}^M\epsilon_{0}(1+\epsilon_{s})}{P_s^MP_0}.
	\end{align}
\end{Corollary}
\begin{IEEEproof}
	By applying Taylor expansion $1-e^{-x}\approx x$ $(x\rightarrow 0)$, the expression in (\ref{Pout_F1}) can be further approximated as follows:
		\begin{align}
	\hat{P}_{out}\approx& \alpha_{1}+\alpha_{s}^M-\alpha_{s}^M\alpha_{1} \notag\\
	    =&\frac{\epsilon_{0}(1+\epsilon_{s})}{P_0}+\frac{\epsilon_{s}^M}{P_s^M}-\frac{\epsilon_{s}^M\epsilon_{0}(1+\epsilon_{s})}{P_s^MP_0},
\end{align}
and the proof is complete.
\end{IEEEproof}

\textbf{\textit{Remark 3.} }From Corollary $3$, it can be easily observed that the proposed FSIC-PA scheme can also avoid outage probability error floors without any constraints on the users' target rates. At this point, the second question raised in Section I can be answered with the answer that HSIC is not the necessary condition to avoid outage probability error floors.

\textbf{\textit{Remark 4.} }It is also interesting to investigate the diversity gain achieved by the FSIC-PA scheme, which is defined as:
\begin{align}
 \hat{d}=-\lim\limits_{P_s\rightarrow\infty}\frac{\log \hat{P}_{out}}{\log P_s}.
\end{align}
According to Corollary $3$, it can be straightforwardly obtained that $\hat{d}=1$. Thus,
the multi-user diversity gain cannot be obtained by FSIC-PA scheme.

The above two remarks indicate that even though HSIC is not the necessary strategy to avoid the outage probability error floor, its combination with PA is beneficial for improving transmission robustness.

\subsection{Comparisons between HSIC-PA/FSIC-PA scheme with HSIC-NPA scheme}
In this section, more detailed comparisons of the proposed two schemes with the benchmark HSIC-NPA scheme are provided. Note that, if the served secondary user belongs to type I, the three schemes, i.e., HSIC-PA, HSIC-NPA and FSIC-PA, achieve the same instantaneous data rate.
However, the three schemes differ from each other if the served secondary user belongs to type II. Thus, it is necessary to compare the three schemes for the case when the served secondary user belongs to type II.
For ease of notation, denote the served secondary user by $U_{m^*}$. When $U_{m^*}$ belongs to type II, denote its achievable rate by ${R}_{II}$, $\hat{R}_{II}$ and $\bar{R}_{II}$ for HSIC-PA, FSIC-PA and HSIC-NPA schemes, respectively.

From the description in Section. II, it can be found that ${R}_{II}\geq\bar{R}_{II}$ always holds. Thus, it is sufficient to characterize the probability of the event that ${R}_{II}>\bar{R}_{II}$, for the comparison between HSIC-PA and HSIC-NPA, as presented in the following theorem.

\begin{Theorem}
Under the condition that the served secondary user $U_{m^*}$ is type II, the probability of the event that ${R}_{II}>\bar{R}_{II}$, termed ${P}^{better}$, is given by:
\begin{align}\label{fenzifenmu}
	{P}^{better}= \frac{P\left( \bar{R}_{\uppercase\expandafter{\romannumeral2}}<R_{\uppercase\expandafter{\romannumeral2}}, U_{m^*}\text{ is type II}\right) }{P\left(U_{m^*}\text{ is type II} \right) },
\end{align}	
where
\begin{align}\label{numerator}
	&P\left( \bar{R}_{\uppercase\expandafter{\romannumeral2}}<R_{\uppercase\expandafter{\romannumeral2}}, U_{m^*}\text{ is type II}\right)\notag\\ =&\sum_{i=1}^{M}\left(\begin{array}{c}
		M \\
		i
	\end{array}\right)(-1)^ie^{\frac{i}{P_s}}\left[\tilde{v}(\alpha_{0},\frac{iP_0}{P_s\alpha_{0}},\frac{i}{P_s}(\alpha_{0}^{-1}-P_0)+1) -\tilde{u}(\alpha_{0},\frac{i}{P_s\alpha_{0}} )  \right] ,
\end{align}
and
\begin{align}
	P\left(U_{m^*}\text{ is type II} \right)=1-\sum_{i=0}^{M}\left(\begin{array}{c}
		M \\
		i
	\end{array}\right)(-1)^ie^{\frac{i}{P_s}}\tilde{u}(\alpha_{0},\frac{i}{P_s\alpha_{0}} ),
\end{align}	
where  {\color{black}
$\tilde{u}(x,y)=\frac{1}{y+1}e^{-x(y+1)}$,
and $\tilde{v}(x,y,z)=\frac{\sqrt{\pi}e^{\frac{z^2}{4y}}}{2\sqrt{y}}[1-\mathrm{erf} (\sqrt{y}(x+\frac{z}{2y}\!))]$, where $\mathrm{erf}(\cdot)$ denotes the Gaussian error function,
which is given by:
\begin{align}
\mathrm{erf}(x)=\frac{2}{\sqrt{\pi}}\int_{0}^{x}e^{-t^2}dt.
\end{align}}
\end{Theorem}
\begin{IEEEproof}
Please refer to Appendix E.
\end{IEEEproof}

Differently, for the comparison between FSIC-PA and HSIC-NPA, $\hat{R}_{II}$ can be either larger or less than $\bar{R}_{II}$. Thus, it is necessary to characterize both the probabilities of the events that $\hat{R}_{II}>\bar{R}_{II}$ and $\hat{R}_{II}<\bar{R}_{II}$. By noting that
	\begin{align}
	\hat{P}\left( \bar{R}_{\uppercase\expandafter{\romannumeral2}}<\hat{R}_{\uppercase\expandafter{\romannumeral2}}, U_{m^*}\text{ is type II}\right)=P\left(|h_M|^2>\frac{\tau(g)}{P_s},|h_M|^2<\left |\overline{h_{k}}\right |^2,|g|^2>\alpha_{0} \right),
	\end{align}
which is the same as the expression of $P\left( \bar{R}_{\uppercase\expandafter{\romannumeral2}}< {R}_{\uppercase\expandafter{\romannumeral2}}, U_{m^*}\text{ is type II}\right)$ in Theorem $3$, the following theorem can be straightforwardly obtained.

\begin{Theorem}
	Under the condition that the served secondary user $U_{m^*}$ is type II, the probability of the event that $\hat{R}_{II}>\bar{R}_{II}$, termed $\hat{P}^{better}$, is given by:
	\begin{align}\label{fenzifenmu}
		\hat{P}^{better}= \frac{P\left( \bar{R}_{\uppercase\expandafter{\romannumeral2}}<\hat{R}_{\uppercase\expandafter{\romannumeral2}}, U_{m^*}\text{ is type II}\right) }{P\left(U_{m^*}\text{ is type II} \right) },
	\end{align}	
which is the same as the expression of ${P}^{better}$ in Theorem $3$.  The probability of the event that $\hat{R}_{II}<\bar{R}_{II}$, termed $\hat{P}^{worse}$,  is given by:
	\begin{align}
	\hat{P}^{worse}=1-\hat{P}^{better}.
		\end{align}
\end{Theorem}

\section{Numerical Results}
In this section, simulation results are provided to verify the accuracy of the developed analysis and demonstrate the performance of the proposed HSIC-PA and FSIC-PA schemes. Comparisons with the benchmark HSIC-NPA scheme developed in \cite{ding2021new,ding2020unveiling1,ding2020unveiling2} are also provided.

\begin{figure}[!t]
	\centering
	\vspace{0em}
	\setlength{\abovecaptionskip}{0em}   
	\setlength{\belowcaptionskip}{-2em}   
	\includegraphics[width=0.6\linewidth]{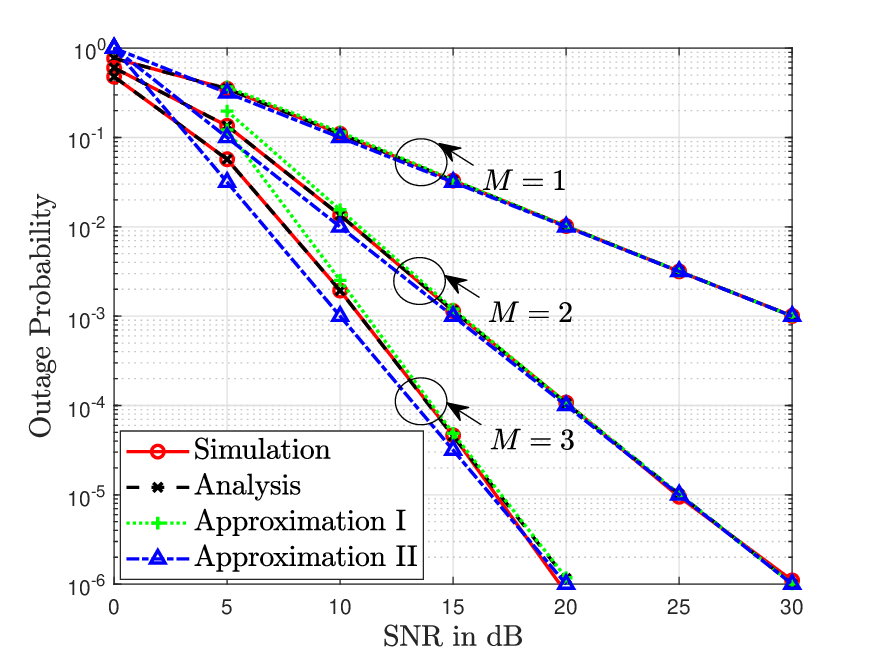}\\
	\caption{Outage probabilities achieved by HSIC-PA scheme. $P_{s}=P_{0}$, $R_{s}=1$ bits per channel use (BPCU) and $R_{0}=1$ BPCU.}
	\label{P_outLFJ_H}
\end{figure}
Fig. \ref{P_outLFJ_H} verifies the accuracy of the developed analytical results for the outage probability achieved by the proposed HSIC-PA scheme. Note that, the curves for analytical results are based on Theorem $1$, and those for Approximations I and II are based on Corollaries $1$ and $2$, respectively.
As shown in the figure, analytical results perfectly match simulations, which verifies the accuracy of the analytical results provided in Theorem $1$.
Besides, Fig. \ref{P_outLFJ_H} also shows that both the curves for Approximation I and Approximation II
match the simulation results at high SNR, which verifies the accuracy of the approximations in Corollaries $1$ and $2$.

\begin{figure}[t]
	\centering
	\vspace{0em}
	\setlength{\abovecaptionskip}{0em}   
	\setlength{\belowcaptionskip}{-2em}   
	\includegraphics[width=0.6\linewidth]{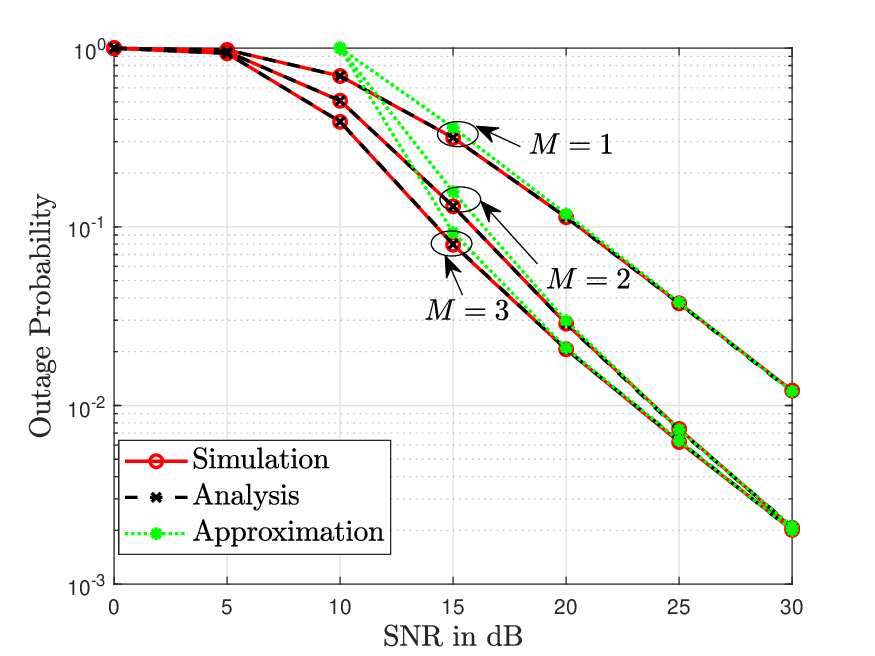}\\
	\caption{Accuracy of the developed analysis and approximation expression of the outage probability achieved by the FSIC-PA scheme. $P_{s}=\frac{P_{0}}{10}$, $R_{s}=1$ BPCU and $R_{0}=1$ BPCU.}
	\label{P_outLFJ_F}
\end{figure}
Fig. \refeq{P_outLFJ_F} verifies the accuracy of the developed analytical results for the outage probability achieved by the proposed FSIC-PA scheme. Note that the curves for analytical results are based on Theorem $2$, and  the curves for approximation are based on Corollary $3$. From the figure, it can be observed that the curves for analysis perfectly match simulations, which verify the accuracy of the results provided in Theorem $2$.
Besides, it is shown that the curves for the approximate results are accurate at high SNR, which demonstrates the accuracy of the results  in Corollary $3$.

A significant  difference between HSIC-PA and FSIC-PA schemes can be clearly observed from Figs. \ref{P_outLFJ_H} and \ref{P_outLFJ_F}. Fig. \ref{P_outLFJ_H} shows that as $M$ increases, the outage probability achieved by HSIC-PA scheme significantly decreases. In contrast, Fig. \ref{P_outLFJ_F} shows that,
for $M>1$, the outage probabilities for different values of $M$ coincide. Thus, keeping increasing $M$ cannot improve the outage performance of FSIC-PA in the high SNR regime.  This observation is consistent with
the results in Section III that the diversity gain of HSIC-PA scheme is $M$, while that of FSIC-PA scheme is only $1$.

\begin{figure}[!t]
	\centering
	\vspace{0em}
	\setlength{\abovecaptionskip}{0em}   
	\setlength{\belowcaptionskip}{-2em}   
	\includegraphics[width=0.6\linewidth]{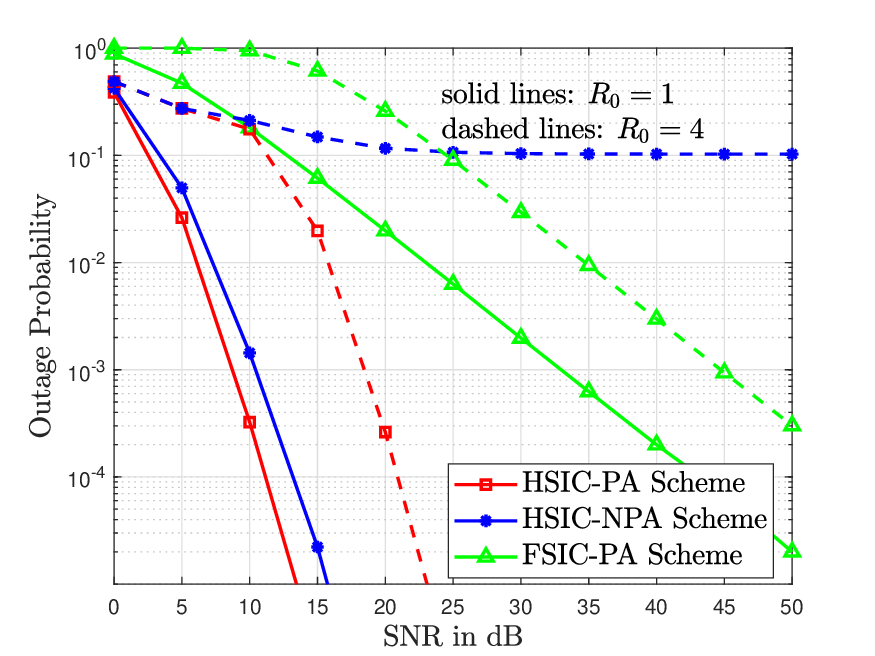}\\
	\caption{Outage probabilities achieved by the HSIC-NPA, HSIC-PA and FSIC-PA schemes versus SNR, $M=4$, $R_{s}=1$ BPCU, $P_{s}=P_{0}$.}
	\label{outage_probability}
\end{figure}
Fig. \ref{outage_probability} shows the outage probabilities of the secondary users achieved by HSIC-NPA, HSIC-PA and FSIC-PA versus transmit SNR. As shown in the figure, for HSIC-NPA scheme, when $R_0=1$ BPCU, there is no outage probability error floor. However, when  $R_0=4$ BPCU, the outage probability error floor exists. This observation is consistent with the conclusions in \cite{ding2021new},
 i.e., the error floor can only be avoided when $\epsilon_0\epsilon_s<1$. By contrast, the proposed HSIC-PA and FSIC-PA schemes can avoid outage probability error floors, since the outage probabilities achieved by both schemes continuously decrease as the SNR increases. Fig. \ref{outage_probability} also shows that the HSIC-PA scheme performs the best among the three schemes for all cases. However, FSIC-PA achieves larger outage probabilities than HSIC-NPA when $R_0=1$ BPCU, while for the case where $R_0=4$ BPCU, FSIC-PA performs better at high SNRs.

\begin{figure}[!t]
	\centering
	\vspace{0em}
	\setlength{\abovecaptionskip}{0em}   
	\setlength{\belowcaptionskip}{-2em}   
	\includegraphics[width=0.6\linewidth]{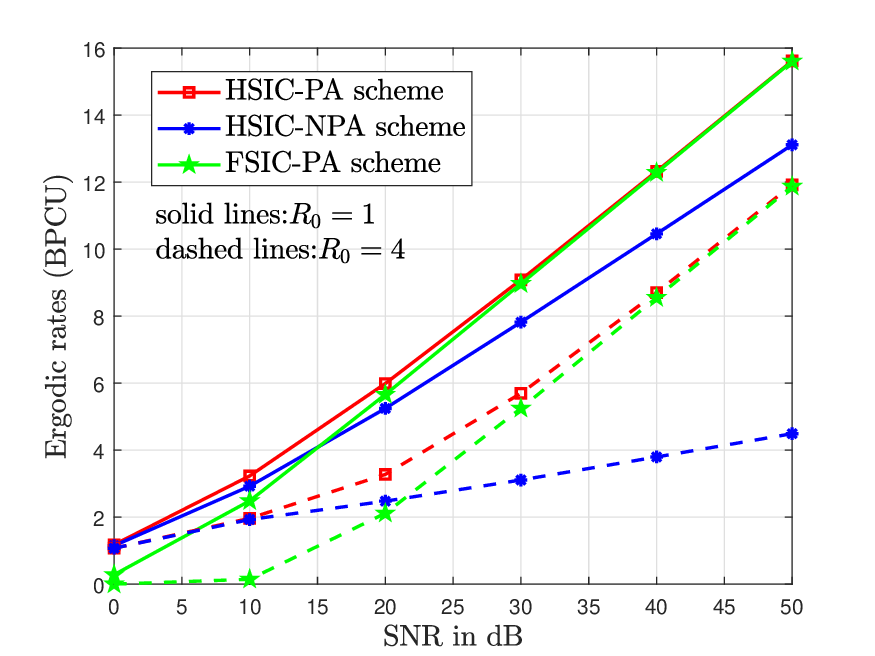}\\
	\caption{Comparison among the three considered transmission schemes in terms of ergodic rates. $M=4$, $P_{s}=P_{0}$.}
	\label{ergodic_rate}
\end{figure}

 Fig. \ref{ergodic_rate} shows the performance of the three schemes in terms of ergodic data rates achieved by the served secondary users.
 From the figure, it is shown that HSIC-PA scheme always achieves the largest ergodic rate among the three schemes, which is consistent with the observation in Fig. \ref{outage_probability}.
 Another interesting observation from  Fig. \ref{ergodic_rate} is that the performance of FSIC-PA approaches that of HSIC-PA in terms of ergodic data rate at high SNR, while the performance of HSIC-NPA approaches that of HSIC-PA in terms of ergodic rate at low SNRs. This observation indicates that it is preferable to set the
 secondary user at the first stage of SIC and use full transmit power at low SNRs, while it is preferable to set the secondary user at the second stage of SIC and use partial transmit power at high SNRs.

\begin{figure}[t]
	\centering
	\vspace{0em}
	\setlength{\abovecaptionskip}{0em}   
	\setlength{\belowcaptionskip}{-2em}   
	\includegraphics[width=0.6\linewidth]{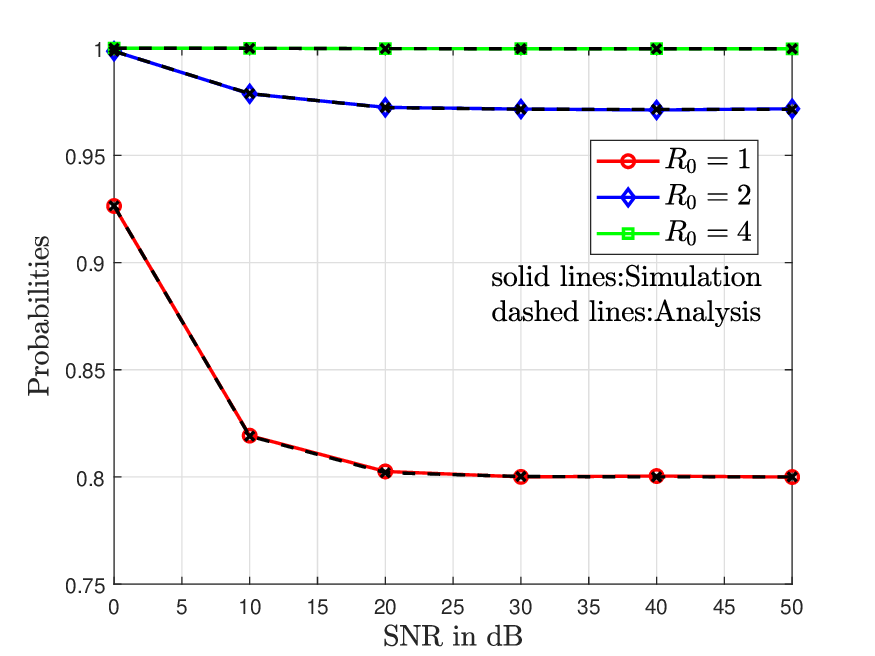}\\
	\caption{The probabilities that the served secondary user belongs to type II. $P_{s}=P_{0}$, $M=4$.}
	\label{P_typeIILF}
\end{figure}

\begin{figure}[t]
	\centering
	\vspace{0em}
	\setlength{\abovecaptionskip}{0em}   
	\setlength{\belowcaptionskip}{-1em}   
	\includegraphics[width=0.6\linewidth]{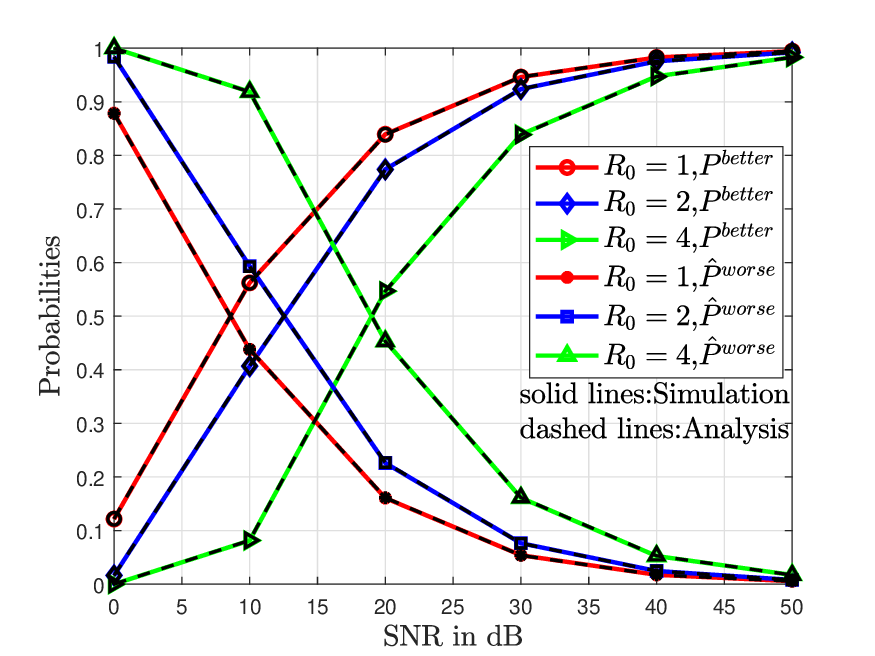}\\
	\caption{$P^{better}$ and $\hat{P}^{worse}$. $P_{s}=P_{0}$, $ M=4 $.}
	\label{P_betterLF}
\end{figure}
Fig. \ref{P_typeIILF} and Fig. \ref{P_betterLF} demonstrate a more detailed comparison on achievable rates of the proposed two schemes with the benchmark HSIC-NPA scheme.
Fig. \ref{P_typeIILF} shows the probability that the served secondary user belongs to type II. It is shown that as SNR increases, the probabilities converge to a constant.
Fig. \ref{P_betterLF} shows that the curves for $\hat{P}^{better}$ and ${P}^{better}$ coincide, which is consistent with results shown in Theorems $3$ and $4$.
Fig. \ref{P_betterLF} also shows that $\hat{P}^{better}$ and ${P}^{better}$ increase with SNR, and approach $1$ in the high SNR regime. While
$\hat{P}^{worse}$ decreases with SNR and approaches $1$ in the low SNR regime.
The above observation can help to understand the phenomenon shown in Fig. \ref{outage_probability} and
Fig. \ref{ergodic_rate}, and leads to the following suggestions for practical systems.
On the one hand, at high SNR, it is preferable to apply power adaptation and put the secondary user at the second stage of SIC. On the other hand, at low SNR, it is better to decode the secondary user at the first stage of SIC.

\begin{figure}[!t]
	\centering
	\vspace{0em}
	\setlength{\abovecaptionskip}{0em}   
	\setlength{\belowcaptionskip}{-2em}   
	\includegraphics[width=0.6\linewidth]{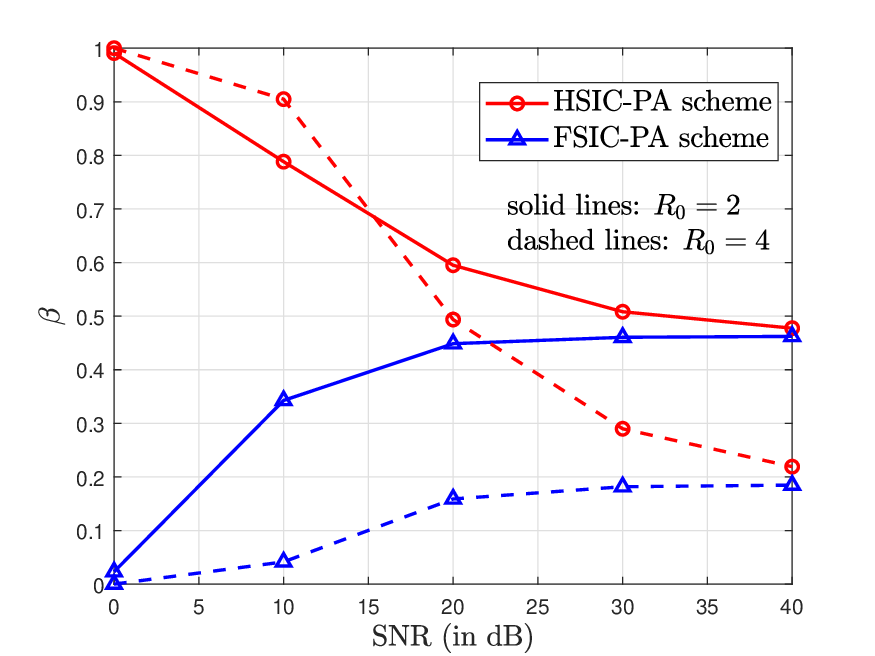}\\
	\caption{Power consumption of the HSIC-PA and FSIC-PA schemes. $ M=4 $, $R_{s}=1$ BPCU, $P_0=P_s$.}
	\label{power_consumption}
\end{figure}
Fig. \ref{power_consumption} shows the power consumption of HSIC-PA and FSIC-PA schemes. Note that  the HSIC-NPA scheme always chooses full power to transmit for the secondary users, i.e., $\beta$ is always set to be $1$, while $\beta$ can be set to be less than $1$ in the proposed HSIC-PA and FSIC-PA schemes. Thus, HSIC-NPA is more energy consuming than the proposed two schemes in this paper. From the figure, it can be observed that at low SNRs, $\beta$ approaches $1$ in HSIC-PA  and $\beta$ approaches zero in FSIC-PA. Besides, as SNR increases, $\beta$ decreases in HSIC-PA, while that in FSIC-PA  increases. More interestingly, the values of $\beta$ for both schemes approach a constant in the high SNR regime. However, at high SNR, the value of $\beta$ in HSIC-PA scheme is a bit higher than that in FSIC-PA.

\section{Conclusions}
In this paper, two novel cognitive radio inspired uplink NOMA schemes were proposed to improve transmission robustness, namely HSIC-PA scheme and FSIC-PA scheme. Rigorous analysis has been developed to characterize the performance of the proposed schemes. It has been shown that both HSIC-PA and FSIC-PA schemes can avoid outage probability error floors in the high SNR regime without any constraints on users' target rates, which was thought impossible for uplink NOMA transmission. It has also been shown that the diversity gain achieved by the HSIC-PA scheme is $M$, which is the maximal multi-user diversity gain for the considered scenario. While the diversity gain achieved by the FSIC-PA scheme is $1$. Numerical results have been presented to verify the accuracy of the developed analysis and demonstrate the superior performance of the proposed schemes. It has been shown by this paper that the combination of HSIC and PA is important to improve the transmission robustness of uplink NOMA.

\appendices
\section{Proof for Lemma 1}
The outage events can be divided into two groups, one is $|g|^2>\alpha_{0}$ and the other is $|g|^2<\alpha_{0}$.
Thus, the outage probability $P_{out}$ shown in (\ref{P_out}) can be written as:
\begin{align}
	P_{out}=&\sum_{m=1}^{M-1} P \left ( E_{m},\max\left\{R^k_I, 1 \le k\le m\right\}<R_s,
	\max\left \{ R^k_{II},m < k  \le M \right \} <R_{s},\left | g \right | ^2>\alpha _{0} \right )\notag\\
	&+P\left( E_{M},\max\left \{ R^k_{I},1\le k  \le M \right \} <R_{s},\left | g \right | ^2>\alpha _{0}  \right)\notag\\
	&+P\left( E_{0},\max\left \{ R^k_{II},1\le k  \le M \right \} <R_{s},\left | g \right | ^2>\alpha _{0}  \right)\notag\\
	&+P\left(\max\left \{ R^k_{II},1\le k  \le M \right \} <R_{s} ,|g|^2<\alpha_{0} \right).
\end{align}

Recall that the secondary users are ordered according to their channel gains, $P_{out}$ can be further written as:
\begin{align}
	P_{out}=&\underbrace{\sum_{m=1}^{M-1} P\left ( E_{m},R^m_I< R_{s},{\color{black}R^M_{II}}  < R_{s},\left | g \right | ^2\!\!>\!\!\alpha _{0} \!\right )}_{Q_{m}} \notag \\
	& +\underbrace{P\left( E_{M},R^M_I<R_{s},\left | g \right | ^2>\alpha _{0}  \right)}_{Q_{M}} +\underbrace{P\left( E_{0},{\color{black}R^M_{II}} <R_{s},\left | g \right | ^2>\alpha _{0}  \right) }_{Q_{0}}\notag \\
	& +\underbrace{ P\left(
		 R^M_{II}<R_{s} ,|g|^2<\alpha_{0} \right) }_{Q_{M+1}}.
\end{align}

{\color{black}Note that when $|g|^2>\alpha _{0}$, $R^M_{II}$ can be determined according to the value of $\left |h_{M} \right |^2$ as follows:}
\begin{align}\label{R^M_{II}}
	R^M_{II}
	=\begin{cases}
		R^M_{II,2}	,&\left |h_{M} \right |^2<{\color{black}\left |\overline{h}\right |^2}\\
		R^M_{II,1}	,&\left |h_{M} \right |^2>{\color{black}\left |\overline{h}\right |^2},
	\end{cases}
\end{align}
where $ {\color{black}\left |\overline{h}\right |^2}=\frac{(\left | g \right |^2\alpha_{0}^{-1}-1 )(P_{0}\left | g \right |^2+1)}{P_{s}}$. Thus{\color{black},} $Q_{m}$ can be rewritten as follows:
\begin{align}
	Q_{m}=&\sum_{m=1}^{M-1}  \underbrace{P\left (E_m,R^m_I<R_{s},R^M_{II,2}<R_{s}, \left | h_{M} \right |^2<{\color{black}\left |\overline{h}\right |^2}  ,|g|^{2}>\alpha_{0} \right ) }_{Q_{m,1}} \notag \\
	&+\underbrace{P\left ( E_m, R^m_I<R_{s},R^M_{II,1}<R_{s},
		\left | h_{M} \right |^2>{\color{black}\left |\overline{h}\right |^2} ,|g|^{2}>\alpha_{0} \right ) }_{Q_{m,2}}.
\end{align}
{\color{black}By noting that regardless of the value of $\left |h_{M} \right |^2$, $R^m_I$ is always smaller than $R^M_{II,1}$ and $R^M_{II,2}$,}
 $Q_m$ can be further simplified {\color{black}as:}
\begin{align}\label{Qm12}
		Q_{m}=&\sum_{m=1}^{M-1}  \underbrace{P\left (E_m,R^M_{II,2}<R_{s}, \left | h_{M} \right |^2<{\color{black}\left |\overline{h}\right |^2} ,|g|^{2}>\alpha_{0} \right ) }_{Q_{m,1}} \notag \\
	&+\underbrace{P\left ( E_m,R^M_{II,1}<R_{s},
		\left | h_{M} \right |^2>{\color{black}\left |\overline{h}\right |^2} ,|g|^{2}>\alpha_{0} \right ) }_{Q_{m,2}}.
\end{align}

By {\color{black}applying the results shown in (\ref{R^M_{II}})}, $Q_{0}$ can be rewritten as follows:
\begin{align}\label{Q012}
	Q_0=&\underbrace{ P\left(\! E_{0},\left |h_{M} \right |^2<{\color{black}\left |\overline{h}\right |^2}, R^M_{II,2} <R_{s},\left | g \right | ^2>\alpha _{0}\!\right)}_{Q_{0,1}}\notag\\
	&+ \underbrace{ P\left( E_{0},\left |h_{M} \right |^2>{\color{black}\left |\overline{h}\right |^2}, R^M_{II,1} <R_{s},\left | g \right | ^2>\alpha _{0}  \right)}_{Q_{0,2}}.
\end{align}

{\color{black}Note that, $Q_{m,1}$ and $Q_{0,1}$ can be combined, so as $Q_{m,2}$ and $Q_{0,2}$, thus, the sum of $Q_m$ and $Q_0$ can be simplified as follows:}
	\begin{align}
	Q_m+Q_0=&\underbrace{Q_{m,1}+Q_{0,1}}_{\tilde{Q}_1} +\underbrace{Q_{m,2}+Q_{0,2}}_{\tilde{Q}_2}\notag\\
	=&\underbrace{P\left(|h_M|^2>\frac{\tau(g)}{P_s},\left | h_{M} \right |^2<{\color{black}\left |\overline{h}\right |^2},R^M_{II,2}<R_s,|g|^{2}>\alpha_{0} \right)}_{\tilde{Q}_1} \notag\\
	+&\underbrace{P\left(|h_M|^2>\frac{\tau(g)}{P_s},\left | h_{M} \right |^2>{\color{black}\left |\overline{h}\right |^2},R^M_{II,1}<R_s,|g|^{2}>\alpha_{0} \right)}_{\tilde{Q}_2}.
\end{align}
Therefore, $P_{out}=Q_m+Q_0+Q_M+Q_{M+1}=\tilde{Q}_1+\tilde{Q}_2+Q_M+Q_{M+1}$ and the proof is complete.

\section{Proof for Theorem 1}
	{\color{black}According to Lemma $1$, the evaluation of $P_{out}$ can be divided into four parts: $\tilde{Q}_1$,
		$\tilde{Q}_2$,
		$Q_M$
		and $Q_{M+1}$.}

\subsection{Evaluation of $\tilde{Q}_1$}
Note that $\tilde{Q}_1$ can be expressed as follows:
\begin{align}
	\tilde{Q}_1\!=&P \! \left(\!|h_M|^2>\frac{|g|^2\alpha_{0}^{-1}\!-\!1}{P_s},|h_M|^2<\frac{(|g|^2\alpha_{0}^{-1}\!-\!1)(P_0|g|^2+1)}{P_s},\log(1\!+\!\tau(g))<R_s,|g|^2>\alpha_{0}\! \right) \notag \\
	=&{\color{black}\underset{\alpha_{0}<|g|^2<\alpha_{1}}{{\Large \boldsymbol{ \varepsilon }}} \bigg\{\underbrace{P\left(\frac{|g|^2\alpha_{0}^{-1}-1}{P_s}<|h_M|^2<\frac{(|g|^2\alpha_{0}^{-1}-1)(P_0|g|^2+1)}{P_s} \right)}_{\tilde{S}_1}\bigg\}},
\end{align}
where ${\color{black}{\Large \boldsymbol{ \varepsilon }}}$$ \left \{ * \right \} $ denotes the mathematical expectation.

Note that the users are ordered according to their channel gains, and hence the probability density function (pdf) of $|h_M|^2$ can be expressed as:
\begin{align}\label{hMpdf}
		f_{|h_{M}|^2}(x)=&\frac{M!}{(M-1)!}(1-e^{-x})^{M-1}e^{-x}\notag\\
=&M(1-e^{-x})^{M-1}e^{-x}.
\end{align}
By {\color{black}applying (\ref{hMpdf})}, $\tilde{S}_1$ can be {\color{black}evaluated} as follows:
\begin{align}
	\tilde{S}_1=&\int_{\frac{|g|^2\alpha_{0}^{-1}\!-\!1}{P_s}}^{\frac{(|g|^2\alpha_{0}^{-1}\!-\!1)(P_0|g|^2+1)}{P_s}}{\color{black}f_{|h_{M}|^2}(x)}dx\notag \\
=&\sum_{i=0}^{M}\left(\begin{array}{c}
	M \\
	i
\end{array}\right)(-1)^i\left( e^{-\frac{i(|g|^2\alpha_{0}^{-1}\!-\!1)(P_0|g|^2+1)}{P_s}}-e^{-\frac{i}{P_s}(|g|^2\alpha_{0}^{-1}\!-\!1)}\right).
\end{align}

{\color{black}Further, by noting that $|g|^2$ is exponentially distributed, $\tilde{Q}_1$ can be calculated as:}
\begin{align}\label{Q1}
	\tilde{Q}_1=&{\color{black}\int_{\alpha_{0}}^{\alpha_{1}}\tilde{S}_1e^{-|g|^2}d|g|^2}\notag\\
	=&	\sum_{i=0}^{M}\left(\begin{array}{c}
		M \\
		i
	\end{array}\right)(-1)^i\int_{\alpha_{0}}^{\alpha_{1}}
\left( e^{-\frac{i(x\alpha_{0}^{-1}\!-\!1)(P_0x+1)}{P_s}}-e^{-\frac{i}{P_s}(x\alpha_{0}^{-1}\!-\!1)}\right)e^{-x}dx.
\end{align}
{\color{black}For notational simplicity,}
{\color{black} define $u(\alpha_{0},\alpha_{1},c)$ as:
		\begin{align} \label{gg}
			u(\alpha_{0},\alpha_{1},c)\overset{\bigtriangleup}{=} \int_{\alpha_{0}}^{\alpha_{1}}e^{-(c+1)x}dx= \frac{1}{c+1}[e^{-\alpha_{0}(c+1)}-e^{-\alpha_{1}(c+1)}],
		\end{align}
and $v(\alpha_{1},\alpha_{0},A,B)$ as:
	\begin{align} \label{hh}
		 v(\alpha_{1},\alpha_{0},A,B)\overset{\bigtriangleup}{=}&\int_{\alpha_{0}}^{\alpha_{1}}e^{-(Ax^2+Bx)}dx\notag\\
		=&\frac{\sqrt{\pi}e^{\frac{B^2}{4A}}}{2\sqrt{A}}[\text{erf}(\!\sqrt{A}(\alpha_{1}+\frac{B}{2A}\!))\!-\! \text{erf}(\sqrt{A}(\alpha_{0}+\frac{B}{2A}\!))].
	\end{align}}
	
	{\color{black}By taking (\ref{gg}) and (\ref{hh}) into (\ref{Q1}),  $\tilde{Q}_1$ can be expressed as:}
	\begin{align}\label{Q_1}
		\tilde{Q}_1=\sum_{i=0}^{M}\left(\begin{array}{c}
			M \\
			i
		\end{array}\right)(-1)^ie^{\frac{i}{P_s}}\left[ {\color{black} v}(\alpha_{1},\alpha_{0},\frac{iP_0}{P_s\alpha_{0}},\frac{i}{P_s}(\alpha_{0}^{-1}-P_0)+1)-{\color{black} u}(\alpha_{0},\alpha_{1},\frac{i}{P_s\alpha_{0}})\right].
	\end{align}
\subsection{Evaluation of $\tilde{Q}_2$}
{\color{black}Note} that $Q_2$ can be expressed as follows:	
\begin{align}
	\tilde{Q}_2=&P  \left(|h_M|^2>\frac{|g|^2\alpha_{0}^{-1}\!-\!1}{P_s},|h_M|^2>\frac{(|g|^2\alpha_{0}^{-1}\!-\!1)(P_0|g|^2+1)}{P_s},\right.\notag\\
	&\left.\log(1\!+\!\frac{P_s|h_M|^2}{P_0|g|^2+1})<R_s,|g|^2>\alpha_{0} \right) \notag \\
	\overset{(a)}{=}&{\color{black}\underset{\alpha_{0}<|g|^2<\alpha_{1}}{ {\Large \boldsymbol{ \varepsilon }}}\bigg\{\underbrace{P\left(\frac{(|g|^2\alpha_{0}^{-1}-1)(P_0|g|^2+1)}{P_s}<|h_M|^2<\alpha_{s}(P_0|g|^2+1) \right)}_{\tilde{S}_2} \bigg\},}
\end{align}
where step (a) is obtained by noting {\color{black}the} hidden condition $\frac{(|g|^2\alpha_{0}^{-1}-1)(P_0|g|^2+1)}{P_s}<\alpha_{s}(P_0|g|^2+1)$, which yields $|g|^2<\alpha_{1}$.

By using the pdf {\color{black}of $|h_M|^2$} shown in (\ref{hMpdf}), $\tilde{S}_2$ can be {\color{black}evaluated} as follows:
\begin{align} \tilde{S}_2=&\int_{\frac{(|g|^2\alpha_{0}^{-1}\!-\!1)(P_0|g|^2+1)}{P_s}}^{\alpha_{s}(P_0|g|^2+1)}M(1-e^{-x})^{M-1}e^{-x}dx\notag\\
=&\sum_{i=0}^{M}\left(\begin{array}{c}
	M \\
	i
\end{array}\right)(-1)^i\left( e^{-i\alpha_{s}(P_0|g|^2+1)}-e^{-\frac{i}{P_s}(|g|^2\alpha_{0}^{-1}\!-\!1)(P_0|g|^2+1)}\right).
\end{align}

{\color{black}Further, by averaging with respect to $|g|^2$}, $\tilde{Q}_2$ can be expressed as:
\begin{align}\label{Q2}
   	\tilde{Q}_2=&\sum_{i=0}^{M}\left(\begin{array}{c}
   	M \\
   	i
   \end{array}\right)(-1)^i\int_{\alpha_{0}}^{\alpha_{1}}\left( e^{-i\alpha_{s}(P_0x+1)}-e^{-\frac{i}{P_s}(x\alpha_{0}^{-1}\!-\!1)(P_0x+1)}\right)e^{-x}dx.
\end{align}
	{\color{black}By taking (\ref{gg}) and (\ref{hh}) into (\ref{Q2}),} $\tilde{Q}_2$ can be {\color{black}further expressed} as follows:
\begin{align}\label{Q_2}
	\tilde{Q}_2=\!\sum_{i=0}^{M}\left(\begin{array}{c}\!
		M \\
		i
	\!\end{array}\right)\!(-1)^i \!\left[e^{-i\alpha_{s}}{\color{black}u} (\alpha_{0},\alpha_{1},i\alpha_{s}P_0)\!-\! e^{\frac{i}{P_s}}{\color{black}v} (\alpha_{1},\alpha_{0},\frac{iP_0}{P_s\alpha_{0}},\frac{i}{P_s}(\alpha_{0}^{-1}\!-\! P_0)+1)\right].
\end{align}
\subsection{Evaluation of $Q_M$}
{\color{black}Note} that $Q_{M}$ can be rewritten as follows:
	\begin{align}\label{QM} Q_{M}=&P\left(|h_{M}|^2<\frac{|g|^2\alpha_{0}^{-1}-1}{P_{s}},\log(1+P_{s}|h_{M}|^2)<R_{s},|g|^2>\alpha_{0}\right) \notag\\
	=&P\left(|h_{M}|^2<\min\left\{\frac{|g|^2\alpha_{0}^{-1}-1}{P_{s}},\alpha_{s}\right\},|g|^2>\alpha_{0}\right)\notag\\
	=&{\color{black}\underset{\alpha_{0}<|g|^2<\alpha_{1}}{{\Large \boldsymbol{ \varepsilon }}}\bigg\{ \underbrace{ P\left(|h_{M}|^2<\frac{\alpha_{0}^{-1}|g|^2-1}{P_{s}}\right) }_{S_{M,1}}\bigg\}  +\underset{|g|^2>\alpha_{1}}{{\Large \boldsymbol{ \varepsilon }}}\bigg\{ \underbrace{P\left(|h_{M}|^2<\alpha_{s}\right)}_{ S_{M,2}} \bigg\}},
\end{align}
where the {\color{black}last} step is obtained by dividing the events into two cases, {\color{black}i.e., $|g|^2<\alpha_{1}$ and $|g|^2>\alpha_{1}$.}

By using the pdf {\color{black}of $|h_M|^2$ shown in (\ref{hMpdf}), the expression for  $S_{M,1}$ and  $S_{M,2}$ can be obtained as:
$S_{M,1}=(1-e^{-\frac{\alpha_{0}^{-1}|g|^2-1}{P_{s}}})^M$ and $S_{M,2}=(1-e^{-\alpha_{s}})^{M}$.}

{\color{black}By averaging with respect to $|g|^2$, $Q_{M}$ can be further evaluated as follows:}
	\begin{align}\label{Q_M}
		Q_{M}=&\int_{\alpha_{0}}^{\alpha_{1}}(1-e^{-\frac{\alpha_{0}^{-1}x-1}{P_{s}}})^Me^{-x}dx+\int_{\alpha_{1}}^{\infty}(1-e^{-\alpha_{s}})^{M}e^{-x}dx\notag\\
		=&\int_{\alpha_{0}}^{\alpha_{1}}\sum\limits_{i=0}^{M}\left(\begin{array}{c}
			M \\
			i
		\end{array}\right)(-1)^ie^{\frac{i}{P_{s}}}e^{-\frac{\alpha_{0}^{-1}}{P_{s}}ix}e^{-x}dx+(1-e^{-\alpha_{s}})^{M}e^{-\alpha_{1}}\notag\\
		=&\sum\limits_{i=0}^{M}\left(\begin{array}{c}
			M \\
			i
		\end{array}\right)(-1)^ie^{\frac{i}{P_{s}}}{\color{black}u} (\alpha_{0},\alpha_{1},\frac{i}{\alpha_{0}P_{s}})+(1-e^{-\alpha_{s}})^{M}e^{-\alpha_{1}},
	\end{align}
{\color{black}where the last step is obtained by applying the results shown {\color{black}in (\ref{gg}).}
\subsection{Evaluation of $Q_{M+1}$}
{\color{black}Note} that $Q_{M+1}$ can be expressed as follows:
$$Q_{M+1}=P\left({\color{black}R^M_{II}}<R_{s} ,|g|^2<\alpha_{0} \right).$$
{\color{black}Note that, when} $|g|^2<\alpha_{0}$, $\tau(g)=0$, {\color{black}yielding $R^M_{II}=\log(1+\frac{ P_{s}|h_{M}|^2}{P_{0}|g|^2+1})$.}
Thus, $Q_{M+1}$ can be further {\color{black}expressed} as:
	\begin{align}
		Q_{M+1}	=&P\left(|g|^2<\alpha_{0},\log(1+\frac{ P_{s}|h_{M}|^2}{P_{0}|g|^2+1})<R_{s}  \right)\notag \\
		=&{\color{black}\underset{|g|^2<\alpha_{0}}{{\Large \boldsymbol{ \varepsilon }}}\bigg\{ \underbrace{ P\left( \log(1+\frac{ P_{s}|h_{M}|^2}{P_{0}|g|^2+1})<R_{s}\right)}_{S_{M+1}} \bigg\}}.
		\end{align}

By using the pdf $|h_M|^2$ shown in (\ref{hMpdf}), $S_{M+1}$ can be evaluated as follows:
\begin{align}
	S_{M+1}=&\int_{0}^{\alpha_{s}(P_{0}|g|^2+1)}f_{|h_M|^2}(x)dx\notag\\
=&(1-e^{-\alpha_{s}(P_{0}|g|^2+1)})^{M}.
\end{align}

Further, by averaging with respect to $|g|^2$, $Q_{M+1}$ can be expressed as:
\begin{align}\label{QM+1}
	Q_{M+1}	=&\int_{0}^{\alpha_{0}}(1-e^{-\alpha_{s}(P_{0}x+1)})^{M}e^{-x}dx\notag\\
	=&\sum_{i=0}^{M}\left(\begin{array}{c}
		M \\
		i
	\end{array}\right)(-1)^ie^{-\alpha_{s}i}\frac{1-e^{-(\alpha_{s}P_{0}i+1)\alpha_{0}}}{\alpha_{s}P_{0}i+1},
\end{align}	
where the last step is obtained by applying the binomial expansion. 	

Therefore, the expressions for $\tilde{Q}_{1}$,
$\tilde{Q}_{2}$,
$Q_M$,
and $Q_{M+1}$ are obtained, and the proof is complete.

\section{Proof for Corollary 1}
In order to facilitate a high SNR approximation, $P_{out}$ in (\ref{T1}) can be
	rewritten as follows:
	\begin{align}
		P_{out}=\sum\limits_{i=0}^{M}\left(\begin{array}{c}
			M \\
			i
		\end{array}\right)(-1)^i\int_{0}^{\alpha_{1}}e^{-x}e^{-i\alpha_{s}(P_0x+1)}dx+(1-e^{-\alpha_{s}})^{M}e^{-\alpha_{1}}.
	\end{align}
	By using the fact that
	\begin{align}
		\sum_{i=0}^{M}\left(\begin{array}{c}
			M \\
			i
		\end{array}\right)(-1)^iA^i=(1-A)^M,
	\end{align}
    $P_{out}$ can be further approximated as follows:
	\begin{align}
		P_{out}=&\int_{0}^{\alpha_{1}}e^{-x}(1-e^{-\alpha_{s}(P_0x+1)})^Mdx+(1-e^{-\alpha_{s}})^{M}e^{-\alpha_{1}}\notag\\
		\approx&\int_{0}^{\alpha_{1}}(1-x)\alpha_{s}^M(P_0x+1)^Mdx+\alpha_{s}^{M}(1-\alpha_{1}),
	\end{align}
	where the last step is obtained by applying Taylor seizes $1-e^{-x}\approx x$ when $x\rightarrow 0$.
	A more simplified form of $P_{out}$ can be obtained by applying the binomial expansion:
	\begin{align}\label{App_int}
		P_{out}
		\approx&\alpha_{s}^M\int_{0}^{\alpha_{1}}(1-x)\sum\limits_{i=0}^{M}\left(\begin{array}{c}
			M \\
			i
		\end{array}\right){P_0}^ix^idx+\alpha_{s}^{M}(1-\alpha_{1})\notag\\
		=&\alpha_{s}^M\int_{0}^{\alpha_{1}}\sum\limits_{i=0}^{M}\left(\begin{array}{c}
			M \\
			i
		\end{array}\right){P_0}^i(x^i-x^{i+1})dx+\alpha_{s}^{M}(1-\alpha_{1}).
	\end{align}
By taking integrations in (\ref{App_int}), $P_{out}$ can be further calculated as follows:
	\begin{align}
		P_{out}
		\approx&\alpha_{s}^M\sum\limits_{i=0}^{M}\left(\begin{array}{c}
			M \\
			i
		\end{array}\right){P_0}^i(\frac{\alpha_{1}^{i+1}}{i+1}-\frac{\alpha_{1}^{i+2}}{i+2})+\alpha_{s}^{M}-\alpha_{s}^{M}\alpha_{1}\notag\\
		\overset{(a)}{=}&\frac{\epsilon_{s}^M}{P_s^MP_0}\sum\limits_{i=0}^{M}\left(\begin{array}{c}
			M \\
			i
		\end{array}\right)\frac{\epsilon_{0}^{i+1}(1+\epsilon_{s})^{i+1}}{i+1}-\frac{\epsilon_{s}^M}{P_s^MP_0^2}\sum\limits_{i=0}^{M}\left(\begin{array}{c}
			M \\
			i
		\end{array}\right)\frac{\epsilon_{0}^{i+2}(1+\epsilon_{s})^{i+2}}{i+2}\notag\\
		&+\frac{\epsilon_{s}^M}{P_s^M}-\frac{\epsilon_{s}^M\epsilon_{0}(1+\epsilon_{s})}{P_s^MP_0}\notag\\
		\overset{(b)}{=}&\frac{\epsilon_{s}^M}{P_s^MP_0}\sum\limits_{i=1}^{M}\left(\begin{array}{c}
			M \\
			i
		\end{array}\right)\frac{\epsilon_{0}^{i+1}(1+\epsilon_{s})^{i+1}}{i+1}-\frac{\epsilon_{s}^M}{P_s^MP_0^2}\sum\limits_{i=0}^{M}\left(\begin{array}{c}
			M \\
			i
		\end{array}\right)\frac{\epsilon_{0}^{i+2}(1+\epsilon_{s})^{i+2}}{i+2}+\frac{\epsilon_{s}^M}{P_s^M},
	\end{align}
	where step (b) is obtained by the fact that the first term shown in step (a) is $\frac{\epsilon_{s}^M\epsilon_{0}(1+\epsilon_{s})}{P_s^MP_0}$ when $i=0$, which is exactly the same as the the last term in {\color{black}step (a)}, and thus can be {\color{black}eliminated} .

\section{Proof for Theorem 2}
Divide the outage events into two cases, one being $|g|^2>\alpha_{0}$ and the other being $|g|^2<\alpha_{0}$. . Therefore, the outage probability $\hat{P}_{out}$ shown in (\ref{hatP_out}) can be rewritten as:
\begin{align}
	\hat{P}_{out}=&\sum_{m=1}^{M-1} P \left ( E_{m},\max\left\{\hat{R}^k_I, 1 \le k\le m\right\}<R_s,
	\max\left \{ \hat{R}^k_{II},m < k  \le M \right \} <R_{s},\left | g \right | ^2>\alpha _{0} \right )\notag\\
	&+P\left( E_{M},\max\left \{ \hat{R}^k_{I},1\le k  \le M \right \} <R_{s},\left | g \right | ^2>\alpha _{0}  \right)\notag\\
	&+P\left( E_{0},\max\left \{ \hat{R}^k_{II},1\le k  \le M \right \} <R_{s},\left | g \right | ^2>\alpha _{0}  \right)\notag\\
	&+P\left(\max\left \{\hat{R}^k_{II},1\le k  \le M \right \} <R_{s} ,|g|^2<\alpha_{0} \right).
\end{align}

Recall that the secondary users are ordered according to their channel gains,
$\hat{P}^{out}$ can be further written as:
\begin{align}
\hat{P}_{out}=&\underbrace{\sum_{m=1}^{M-1}P\left(E_m,\hat{R}_{I}^{m}<R_s,\hat{R}_{II}^{M}<R_s,|g|^2>\alpha_0\right)}_{F_m}
+\underbrace{P\left(E_M,\hat{R}_{I}^{M}<R_s,|g|^2>\alpha_0 \right)}_{F_M}\notag\\
+&\underbrace{P\left(E_0,\hat{R}_{II}^{M}<R_s,|g|^2>\alpha_0 \right)}_{F_0}
+\underbrace{P\left(\hat{R}_{II}^{M}<R_s,|g|^2<\alpha_0 \right)}_{F_{M+1}}.
\end{align}

By noting that $\hat{R}^m_I<\hat{R}^M_{II}$ for the first term,  $F_m$ and $F_0$ can be combined as follows:
\begin{align}
F_m+F_0=\underbrace{P\left(|h_{M}|^2>\frac{\tau(g)}{P_{s}},\hat{R}^M_{II}<R_{s},
|g|^2>\alpha_{0}\right)}_{\tilde{F}}.
\end{align}

Therefore, $\hat{P}_{out}$ can be further simplified as:
\begin{align}\label{outage probability of FSIC-PA}
	\hat{P}_{out}=&\underbrace{P\left(|h_{M}|^2<\frac{\tau(g)}{P_{s}},\hat{R}^M_I<R_{s}
,|g|^2>\alpha_{0}\right)}_{F_{M}}
+\underbrace{P\left(\hat{R}_{II}^{M}<R_s,|g|^2<\alpha_{0}\right)}_{F_{M+1}}\notag\\ &+\underbrace{P\left(|h_{M}|^2>\frac{\tau(g)}{P_{s}},
\hat{R}^M_{II}<R_{s},|g|^2>\alpha_{0}\right)}_{\tilde{F}}.
\end{align}

Thus the remaining task is to derive the expressions for $F_M$, $F_{M+1}$ and $\tilde{F}$, respectively.

\subsection{Evaluation of $F_M$}
Note that $F_{M}$ can be expressed as follows:
\begin{align}
F_M=&P\left(|h_M|^2<\frac{|g|^2{\alpha_0}^{-1}-1}{P_s},
\log(1+P_s|h_M|^2)<R_s,|g|^2>\alpha_0 \right),
\end{align}
which is the same as the expression for $Q_M$ in (\ref{QM}). Thus, $F_M$ can be expressed as:
\begin{align}
F_M=\sum\limits_{i=0}^{M}\left(\begin{array}{c}
			M \\
			i
		\end{array}\right)(-1)^ie^{\frac{i}{P_{s}}}{\color{black}u}(\alpha_{0},\alpha_{1},
\frac{i}{\alpha_{0}P_{s}})+(1-e^{-\alpha_{s}})^{M}e^{-\alpha_{1}}.
\end{align}

\subsection{Evaluation of $F_{M+1}$}
Note that $F_{M+1}$ can be expressed as follows:
\begin{align} F_{M+1}=&P\left(\log(1+\tau(g))<R_s,|g|^2<\alpha_{0}\right)\notag\\
\overset{(a)}{=}&P\left(|g|^2<\alpha_{0}\right)\notag\\
=&1-e^{-\alpha_{0}},
\end{align}
where step (a) is obtained by the fact that $\tau(g)=0$ when $|g|^2<\alpha_0$.

\subsection{Evaluation of $\tilde{F}$}
Note that $\tilde{F}$ can be expressed as follows:
\begin{align}\label{F}
\tilde{F}=&P\left(|h_{M}|^2>\frac{\tau(g)}{P_{s}},
\log(1+\tau(g))<R_{s},|g|^2>\alpha_{0}  \right)\notag\\
	=&\underset{\alpha_{0}<|g|^2<\alpha_{1}}{ {\Large \boldsymbol{ \varepsilon }}}\bigg\{\underbrace{P\left(|h_M|^2>\frac{|g|^2{\alpha_0}^{-1}-1}{P_s} \right)}_{\tilde{T}} \bigg\}.
\end{align}

By using the pdf of $|h_M|^2$ shown in (\ref{hMpdf}), $\tilde{T}$ can be evaluated as follows:
\begin{align}\label{T}
\tilde{T}=&\int_{\frac{|g|^2{\alpha_0}^{-1}-1}{P_s}}^{\infty}
M(1-e^{-x})^{M-1}e^{-x}dx\notag\\
=&1-\sum_{i=0}^{M}\left(\begin{array}{c}
		M \\
		i \end{array}\right)(-1)^ie^{-\frac{i}{P_{s}\alpha_{0}}|g|^2+\frac{i}{P_{s}}}.
\end{align}

By taking expectation with respect to $|g|^2$, $\tilde{F}$ can be further evaluated as follows:
\begin{align}
\tilde{F}=&\int_{\alpha_{0}}^{\alpha_{1}}
	\left( 1-\sum_{i=0}^{M}\left(\begin{array}{c}
		M \\
		i
	\end{array}\right)(-1)^ie^{-\frac{i}{P_{s}\alpha_{0}}x+\frac{i}{P_{s}}}\right) e^{-x}dx\notag\\
	=&e^{-\alpha_{0}}-e^{-\alpha_{1}}-\sum_{i=0}^{M}\left(\begin{array}{c}
		M \\
		i
	\end{array}\right)(-1)^ie^{\frac{i}{P_{s}}}{\color{black}u}(\alpha_{0},\alpha_{1},\frac{i}{P_{s}\alpha_{0}}).
\end{align}

Until now, the expressions for $F_M$, $F_{M+1}$ and $\tilde{F}$ are obtained, and the proof is complete.

\section{Proof for Theorem 3}
Note that the numerator in (\ref{fenzifenmu}) can be rewritten as:
\begin{align}
&\underbrace{P\left( \bar{R}_{\uppercase\expandafter{\romannumeral2}}<R_{\uppercase\expandafter{\romannumeral2}}, U_{m^*}\text{ is type II}\right)}_{Q_n}\notag\\
=&\underset{|g|^2>\alpha_{0}}{ {\Large \boldsymbol{ \varepsilon }}}\bigg\{\underbrace{P\left(\frac{|g|^2\alpha_{0}^{-1}-1}{P_s}<|h_M|^2
<\frac{(|g|^2\alpha_{0}^{-1}-1)(P_0|g|^2+1)}{P_s} \right)}_{S_n} \bigg\}.
\end{align}

By using the pdf of $|h_M|^2$ shown in (\ref{hMpdf}), $S_n$ can be evaluated as follows:
\begin{align}
	S_n=&\int_{\frac{|g|^2\alpha_{0}^{-1}\!-\!1}{P_s}}^{\frac{(|g|^2\alpha_{0}^{-1}\!-\!1)(P_0|g|^2+1)}{P_s}}M(1-e^{-x})^{M-1}e^{-x}dx\notag \\
	=&\sum_{i=0}^{M}\left(\begin{array}{c}
		M \\
		i
	\end{array}\right)(-1)^i\left( e^{-\frac{i(|g|^2\alpha_{0}^{-1}\!-\!1)(P_0|g|^2+1)}{P_s}}-e^{-\frac{i}{P_s}(|g|^2\alpha_{0}^{-1}\!-\!1)}\right).
\end{align}

Further, by averaging with respect to $|g|^2$, $Q_n$ can be expressed as follows:
\begin{align}\label{Qn}
	Q_n=&\sum_{i=0}^{M}\left(\begin{array}{c}
		M \\
		i
	\end{array}\right)(-1)^i\int_{\alpha_{0}}^{\infty}
	\left( e^{-\frac{i(x\alpha_{0}^{-1}\!-\!1)(P_0x+1)}{P_s}}-e^{-\frac{i}{P_s}(x\alpha_{0}^{-1}\!-\!1)}\right)e^{-x}dx.
\end{align}

By taking (\ref{gg}) and (\ref{hh}) into (\ref{Qn}), $Q_n$ can be further expressed as follows:
\begin{align}\label{Q_n}
	Q_n=&\sum_{i=0}^{M}\left(\begin{array}{c}
		M \\
		i
	\end{array}\right)(-1)^ie^{\frac{i}{P_s}}\left[ {\color{black}v}(\infty,\alpha_{0},\frac{iP_0}{P_s\alpha_{0}},\frac{i}{P_s}(\alpha_{0}^{-1}-P_0)+1)-{\color{black}u}(\alpha_{0},\infty,\frac{i}{P_s\alpha_{0}})\right]\notag\\
=&\sum_{i=1}^{M}\left(\begin{array}{c}
	M \\
	i
\end{array}\right)(-1)^ie^{\frac{i}{P_s}}\left[ {\color{black} \tilde{v}(\alpha_{0},\frac{iP_0}{P_s\alpha_{0}},\frac{i}{P_s}(\alpha_{0}^{-1}-P_0)+1)
-\tilde{u}(\alpha_{0},\frac{i}{P_s\alpha_{0}})}\right],
\end{align}
where the last step is obtained by noting that the term $i=0$ can be omitted since ${\color{black} \tilde{v}(\alpha_{0},\frac{iP_0}{P_s\alpha_{0}},\frac{i}{P_s}(\alpha_{0}^{-1}-P_0)+1)
-\tilde{u}(\alpha_{0},\frac{i}{P_s\alpha_{0}})}
=0$ for $i=0$.

The denominator in (\ref{fenzifenmu}) can be calculated as follows:
\begin{align}\label{Q_d}
	&\underbrace{P\left(U_{m^*}\text{ is type II} \right)}_{Q_d}\notag\\ =&\underbrace{P\left(|h_M|^2>\frac{\tau(g)}{P_s},|g|^2>\alpha_{0} \right)}_{Q_{d1}}
+\underbrace{P\left( |g|^2<\alpha_{0}\right)}_{Q_{d2}} \notag\\
=&\underset{|g|^2>\alpha_{0}}{ {\Large \boldsymbol{ \varepsilon }}}\bigg\{\underbrace{P\left(
|h_M|^2>\frac{|g|^2\alpha_{0}^{-1}-1}{P_s}
\right)}_{S_{d1}} \bigg\}
+Q_{d2}.
\end{align}

Note that $S_{d1}$ is the same as the expression for $\tilde{T}$ in (\ref{F}). Thus, $S_{d1}$ can be obtained by using the
results in (\ref{T}) as follows:
\begin{align}
S_{d1}=&1-\sum_{i=0}^{M}\left(\begin{array}{c}
		M \\
		i \end{array}\right)(-1)^ie^{-\frac{i}{P_{s}\alpha_{0}}|g|^2+\frac{i}{P_{s}}}.
\end{align}

By averaging with respect to $|g|^2$, $Q_{d1}$ can be further evaluated as follows:
\begin{align}
Q_{d1}=&\int_{\alpha_{0}}^{\infty}
	\left( 1-\sum_{i=0}^{M}\left(\begin{array}{c}
		M \\
		i
	\end{array}\right)(-1)^ie^{-\frac{i}{P_{s}\alpha_{0}}x+\frac{i}{P_{s}}}\right) e^{-x}dx\notag\\
	=&e^{-\alpha_{0}}-\sum_{i=0}^{M}\left(\begin{array}{c}
		M \\
		i
	\end{array}\right)(-1)^ie^{\frac{i}{P_{s}}}{\color{black}\tilde{u}(\alpha_{0},\frac{i}{P_s\alpha_{0}}).}
\end{align}

$Q_{d2}$ can be expressed as follows:
\begin{align}
Q_{d2}=\int_{0}^{\alpha_0}e^{-x}dx=1-e^{-\alpha_0}.
\end{align}

Thus, $Q_d$ is the sum of $Q_{d1}$ and $Q_{d2}$, which can be expressed as follows:
\begin{align}	
Q_d=&1-\sum_{i=0}^{M}\left(\begin{array}{c}
	M \\
	i
\end{array}\right)(-1)^ie^{\frac{i}{P_{s}}}{\color{black}\tilde{u}(\alpha_{0},\frac{i}{P_s\alpha_{0}}).}
\end{align}

Therefore, the expressions for $P\left( \bar{R}_{\uppercase\expandafter{\romannumeral2}}<R_{\uppercase\expandafter{\romannumeral2}}, U_{m^*}\text{ is type II}\right)$ and $P\left(U_{m^*}\text{ is type II} \right)$ are obtained, and the proof is complete.

\bibliographystyle{IEEEtran}
\bibliography{IEEEabrv,ref}
\end{document}